\documentclass[conference]{IEEEtran}
\IEEEoverridecommandlockouts
\usepackage{cite}
\usepackage{amsmath,amssymb,amsfonts}
\usepackage{algorithmic}
\usepackage{graphicx}
\usepackage{textcomp}
\usepackage{xcolor}
\def\BibTeX{{\rm B\kern-.05em{\sc i\kern-.025em b}\kern-.08em
  T\kern-.1667em\lower.7ex\hbox{E}\kern-.125emX}}
\begin{document}

\title{Make It Hard to Hear, Easy to Learn: Long-Form Bengali ASR and Speaker Diarization via Extreme Augmentation and Perfect Alignment
}

\author{
\IEEEauthorblockN{Sanjid Hasan}
\IEEEauthorblockA{\textit{EEE}, \textit{KUET}\\
sanjid9657@gmail.com}
\and
\IEEEauthorblockN{Risalat Labib}
\IEEEauthorblockA{\textit{CSE}, \textit{BUET}\\
risalatlabib@gmail.com}
\and
\IEEEauthorblockN{A H M Fuad}
\IEEEauthorblockA{\textit{CSE}, \textit{BUET}\\
ahmfuad9@gmail.com}
\and
\IEEEauthorblockN{Bayazid Hasan}
\IEEEauthorblockA{\textit{CSE}, \textit{BUET}\\
bhsarf@gmail.com}
}

\maketitle

\begin{abstract}
Although Automatic Speech Recognition (ASR) in Bengali has seen significant progress, processing long-duration audio and performing robust speaker diarization remain critical research gaps. To address the severe scarcity of joint ASR and diarization resources for this language, we introduce Lipi-Ghor-882, a comprehensive 882-hour multi-speaker Bengali dataset. In this paper, detailing our submission to the DL Sprint 4.0 competition, we systematically evaluate various architectures and approaches for long-form Bengali speech. For ASR, we demonstrate that raw data scaling is ineffective; instead, targeted fine-tuning utilizing perfectly aligned annotations paired with synthetic acoustic degradation (noise and reverberation) emerges as the singular most effective approach. Conversely, for speaker diarization, we observed that global open-source state-of-the-art models (such as Diarizen) performed surprisingly poorly on this complex dataset. Extensive model retraining yielded negligible improvements; instead, strategic, heuristic post-processing of baseline model outputs proved to be the primary driver for increasing accuracy. Ultimately, this work outlines a highly optimized dual pipeline achieving a $\sim$0.019 Real-Time Factor (RTF), establishing a practical, empirically backed benchmark for low-resource, long-form speech processing.
\end{abstract}


% \section{Introduction}
% The transcription and segmentation of multi-speaker Bengali audio is a highly complex task, historically constrained by a lack of large-scale, temporally aligned conversational datasets \cite{b1}. In real-world applications and competitions such as DL Sprint 4.0 (BUET CSE FEST 2026), models are evaluated on long-duration audio (e.g., hours-long test sets). Under these conditions, standard sequence-to-sequence models suffer from hallucination, and diarization systems fail to track speaker turn-taking accurately amidst background noise and dialectal shifts.

% In this paper, we detail the solution developed by Team Villagers. Our core observation is twofold: 
% 1) \textbf{For ASR}, feeding clean data to models fails to generalize. Robustness is only achieved by introducing heavy acoustic corruption (20\% noise augmentation) while maintaining immaculate text alignment---a paradigm we term ``Make It Hard to Hear, Easy to Learn.''
% 2) \textbf{For Speaker Diarization}, global state-of-the-art (SOTA) architectures fail to generalize to the acoustic realities of this competition. Heavy retraining yields no measurable benefit, whereas strict algorithmic post-processing of baseline segmentations drastically improves the Diarization Error Rate (DER).

% To support this research, we introduce the \textbf{Lipi-Ghor-882} corpus, an 882-hour dataset bridging the gap between ASR and diarization tasks.

\section{Introduction}
The transcription and segmentation of multi-speaker Bengali audio is a highly complex task, historically constrained by a lack of large-scale, temporally aligned conversational datasets. In the DL Sprint 4.0 (BUET CSE FEST 2026) competition, models were rigorously evaluated on a massive 22-hour hidden test set consisting of long-form audio. This paper details the journey of Team Villagers a process of exhaustive trial and error that forced us to discard conventional wisdom, leading to our core methodology: \textit{Make it hard to hear, easy to learn.}

Our journey began with a straightforward evaluation of pre-trained Bengali Automatic Speech Recognition (ASR) models. We tested Moonshine, Hishab-Titu-Bn-Conformer-Large, Wav2Vec2, and Whisper-Medium (fine-tuned by Tugstugi). The initial results presented a stark trade-off between speed and accuracy. Moonshine was blisteringly fast (5 minutes for the 22-hour set) but performed very poorly. Titu-FastConformer was exceptionally fast (17 minutes) and competitive. Wav2Vec2 took 2 hours with chunking, while Whisper-Medium delivered the best accuracy but took an agonizing 4 hours to run.

Because the Real-Time Factor (RTF) and inference time were critical competition metrics, our next phase was pure optimization. We aggressively optimized Wav2Vec2 with batching to drop its time to 1 hour, but Titu-FastConformer remained superior in speed. However, because Whisper-Medium held the highest accuracy potential, we focused our engineering efforts there. Transitioning to Kaggle's dual T4 environment, we utilized CTranslate2 (CT2) and \texttt{faster-whisper} with parallel processing. To handle long-form context, we experimented with Silero VAD and Pyannote VAD for chunking; Silero VAD proved far more reliable. This engineering phase was a massive success, dropping Whisper's inference time from 4 hours to just 26 minutes (an RTF of $\sim$0.019). We also discovered that \texttt{faster-whisper} outperformed \texttt{whisperX} for our use case, though we had to manually adjust the default VAD parameters, which were initially truncating the first words of speech chunks.

With inference optimized, we entered the training phase, which quickly became a graveyard of failed hypotheses. We attempted Parameter-Efficient Fine-Tuning (PEFT) with a frozen decoder, which yielded negligible gains. We integrated Contextual Biasing \cite{Lall_2024}, which initially improved our scores but began degrading performance catastrophically after a certain training threshold, completely failing on out-of-distribution data. We tried applying Demucs for music and noise removal; while it improved the ASR score, the computational overhead inflated inference time beyond acceptable limits, forcing us to abandon it. Finally, we attempted to ensemble all our best model outputs using ROVER, but to our surprise, the ensembled predictions performed worse than our standalone models.

Our breakthrough in ASR finally came when we shifted our focus from the models to the data. We trained Whisper on a small, perfectly aligned subset of data where 20\% of the audio was artificially corrupted with random noise and reverberation. This counter-intuitive approach forced the model to rely on deep phonetic features rather than acoustic memorization, leading to a significant score boost. We continued this training until we hit a ceiling caused by the poor annotation quality of the broader dataset. 

Simultaneously, our speaker diarization journey was fraught with similar challenges. We tested a suite of models including ECAPA-TDNN, Pyannote 3.1, Pyannote Community-1, and Diarizen. Surprisingly, Diarizen-the supposed open-source state-of-the-art performed poorly on both the public and private leaderboards. Integrating a Bengali-trained WavLM with Diarizen or using VBx with a Pyannote pipeline yielded no changes. In stark contrast to ASR, using Demucs for noise removal actually worsened our diarization scores. Hoping to force robustness, we trained the segmentation model on muffled audio for 20 epochs, but observed negligible improvement in the Diarization Error Rate (DER). 

Ultimately, we realized that for Bengali diarization, model retraining was a dead end. Success relied entirely on heuristic post-processing. While Pyannote 3.1 topped the public leaderboard, Pyannote Community-1 proved more robust for the private leaderboard. We paired this base model with a strict custom post-processing algorithm that forced inter-speaker gaps, merged same-speaker micro-segments, and ruthlessly mitigated overlaps.

This exhaustive journey from the failure of global SOTAs to the success of noisy training and algorithmic post-processing highlighted a severe lack of joint ASR and diarization data in the Bengali domain. To bridge this gap, and as the final piece of our contribution, we created \textbf{Lipi-Ghor-882}, an 882-hour dataset engineered utilizing \texttt{yt-dlp} and the Pyannote API, providing the foundation for our methodology.

\section{The Lipi-Ghor-882 Dataset}A significant research gap exists in the joint optimization of ASR and diarization for Bengali. To address this, we curated Lipi-Ghor-882. We collected diverse long, medium, and short audio files from YouTube spanning multiple domains. Transcriptions were extracted utilizing the \texttt{yt-dlp} library. To establish speaker boundaries, we utilized the current SOTA Pyannote API, carefully merging these temporal annotations with the speech segments. \cite{lipighor2025}

\begin{table}[htbp]
\caption{Lipi-Ghor Dataset Statistics}
\label{tab:dataset-stats}
\centering
\resizebox{\columnwidth}{!}{%
\begin{tabular}{|l|r|l|r|}
\hline
\textbf{Attribute} & \textbf{Value} & \textbf{Attribute} & \textbf{Value} \\
\hline
Total Hours & $\sim$882 & Unique Channels & 596 \\
\hline
Annotated Hours & $\sim$856 & Categorized Domains & 150+ \\
\hline
Total Videos & 1,019 & Annotation Format & SSTT \\
\hline
\end{tabular}%
}
\end{table}

\begin{figure*}[htbp]
  \centering
  % First sub-figure
  \begin{minipage}[b]{0.48\textwidth}
    \centering
    \includegraphics[width=\textwidth]{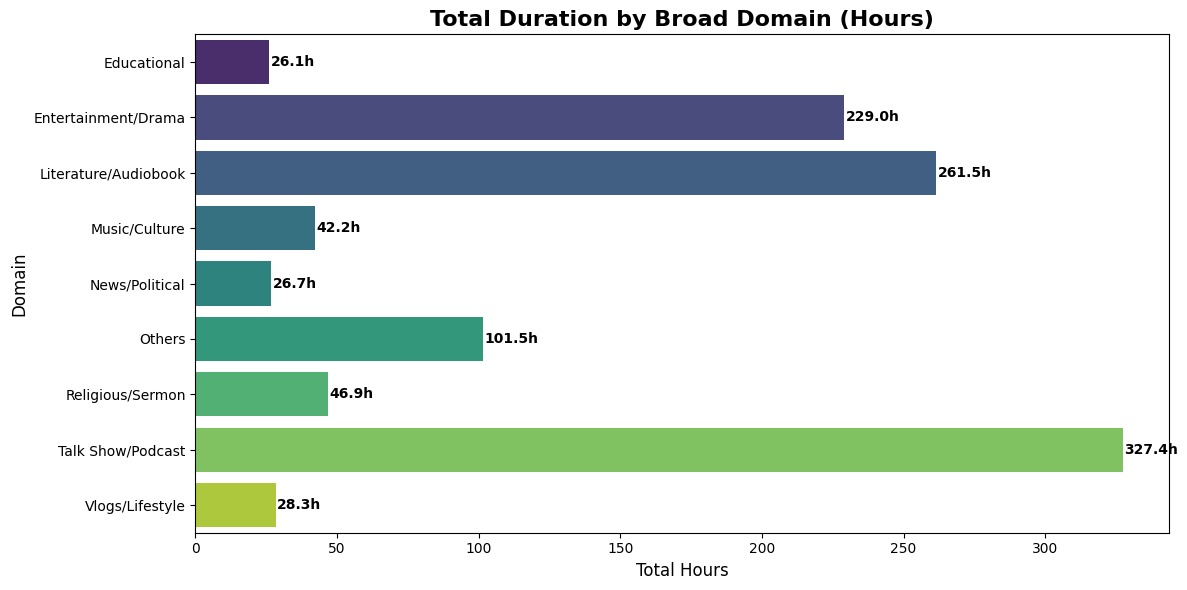}
    % \caption{Total Duration of Audios by Domain.}
    \label{fig:image1}
  \end{minipage}
  \hfill % This adds flexible space between the two images
  % Second sub-figure
  \begin{minipage}[b]{0.48\textwidth}
    \centering
    \includegraphics[width=\textwidth]{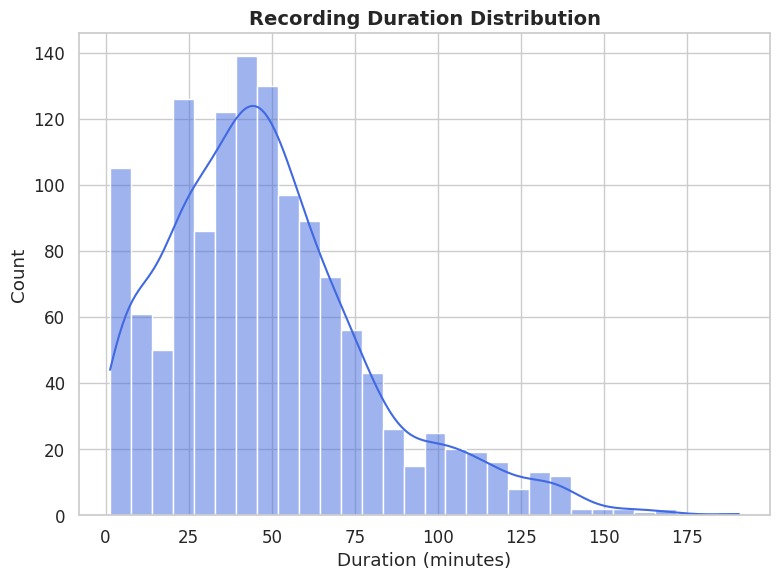}
    % \caption{This is the caption for the second image.}
    \label{fig:image2}
  \end{minipage}
  
  % Optional: A main caption for the entire full-width block
  \caption{Duration Classification of Audios}
  \label{fig:overall_eda}
\end{figure*}

\section{Methodology: Automatic Speech Recognition}

\subsection{Base Model Evaluation and Inference Profiling}
We initially evaluated multiple pre-trained Bengali base models on a 22-hour validation set. Inference time and the Real-Time Factor (RTF) were critical constraints.
\begin{itemize}
  \item \textbf{Moonshine:} Highly efficient (5 minutes inference) but exhibited very poor transcription accuracy.
  \item \textbf{Hishab-Titu-Bn-Conformer-Large:} Exceptionally fast (17 minutes) with competitive accuracy.
  \item \textbf{Wav2Vec2:} Slower (2 hours with chunking) and moderate accuracy.
  \item \textbf{Whisper-Medium (Tugstugi):} Longest base inference (4 hours) but yielded the highest transcription fidelity. 
\end{itemize}
For sequence chunking, we compared Silero VAD and Pyannote VAD; Silero VAD demonstrated superior boundary detection for our specific use case.

\subsection{Optimization and Pipeline Engineering}
Given Whisper's superior accuracy and Titu-Conformer's speed, we focused on optimizing their inference. By converting the Whisper-Medium model to CTranslate2 (CT2) format and deploying it via \texttt{faster-whisper} on dual T4 GPUs (parallel processing), we reduced the 4-hour inference time to \textbf{26 minutes (RTF $\sim$0.019)}. We observed that the default VAD parameters in \texttt{faster-whisper} aggressively truncated initial words in speech chunks; manually fine-tuning these parameters was necessary to prevent word-deletion errors. 

\subsection{Training Ablations and the Noise Paradigm}
Extensive training ablations were conducted to maximize the Whisper-Medium architecture:
\begin{itemize}
  \item \textbf{Frozen Decoder:} Training with a frozen decoder using adapters yielded improvements but not impressive.
  \item \textbf{Contextual Biasing:} Implementing contextual biasing \cite{Lall_2024} improved initial scores but caused catastrophic degradation after a training threshold.
  \item \textbf{Audio Separation (Demucs):} While applying Demucs to remove background music improved the WER, it drastically inflated inference time and was discarded.
  \item \textbf{Ensembling (ROVER):} Attempting to ensemble our best predictions via ROVER degraded the final score. 
  \item \textbf{Final Strategy (Corrupted Audio):} Training the model on a small, perfectly aligned dataset where 20\% of the audio was artificially corrupted (noise and reverberation) forced the model to learn robust acoustic features, leading to our best score. 
\end{itemize}

\section{Methodology: Speaker Diarization}

\subsection{Evaluation of Global SOTAs}
We evaluated several standard and SOTA diarization models, including ECAPA-TDNN, Pyannote 3.1, Pyannote Community 1, and Diarizen. Despite Diarizen being recognized as a leading open-source SOTA \cite{lanzendörfer2025benchmarkingdiarizationmodels}, it performed poorly on both leaderboards for this dataset. Integrating a Bengali-trained WavLM into Diarizen or testing VBx \cite{pálka2025vbxendtoendneuralclusteringbased} with a Pyannote pipeline resulted in no performance shift. Furthermore, unlike in ASR, utilizing Demucs for noise removal worsened the diarization score.

\subsection{The Failure of Retraining}
We attempted to fine-tune the Pyannote segmentation model utilizing artificially muffled data to improve robustness. However, even after 20 epochs, no improvement in DER was observed. 

\subsection{Algorithmic Post-Processing (The Strategy)}
Because model retraining failed, our final solution relied entirely on Pyannote Community-1 paired with a highly structured overlap mitigation and segment merging strategy. We converted our pipeline logic into the following heuristic post-processing algorithm to force strict inter-speaker gaps and filter micro-segments.

\vspace{5pt}
\noindent \textbf{Algorithm 1: Strict Gap Post-Processing}
\begin{enumerate}
  \item \textbf{Input:} Set of detected segments $S$, thresholds $\theta_{merge}=3.79s$, $\theta_{gap}=0.17s$, $\theta_{seg}=0.75s$, $\theta_{spk}=9.0s$.
  \item \textbf{Sort:} Order $S$ chronologically by start time.
  \item \textbf{Strict Rename:} Reassign speaker IDs serially (e.g., SPEAKER\_0, SPEAKER\_1) based on first appearance.
  \item \textbf{Overlap Resolution \& Gap Enforcement:}
  \begin{itemize}
    \item Initialize $timeline\_end = 0.0$, $last\_spk = Null$
    \item For each segment in $S$:
    \item If $start\_time < timeline\_end$, set $start\_time = timeline\_end$.
    \item If $speaker\_id \neq last\_spk$:
    \item $req\_start = timeline\_end + \theta_{gap}$
    \item If $start\_time < req\_start$, set $start\_time = req\_start$.
    \item Update $timeline\_end = end\_time$, $last\_spk = speaker\_id$.
  \end{itemize}
  \item \textbf{Segment Merging:} Merge adjacent segments of the same speaker if the temporal gap is $< \theta_{merge}$.
  \item \textbf{Duration Filtering:} Discard any individual segment where $(end\_time - start\_time) < \theta_{seg}$.
  \item \textbf{Speaker Pruning:} Discard all segments belonging to any speaker with a total speaking duration $< \theta_{spk}$.
  \item \textbf{Output:} Cleaned and rounded temporal segments.
\end{enumerate}

\section{Experiments and Results}

Trainings were done using L40S GPU, 48GB VRAM. All inference processes were evaluated under strict hardware constraints (2x T4 GPUs). The separation of the ASR and Diarization pipelines allowed for maximum computational efficiency without relying on complex, end-to-end joint modeling.

\begin{table}[htbp]
\caption{ASR Performance Benchmarks on 22-Hours Test Set}
\begin{center}
\begin{tabular}{|l|c|c|c|}
\hline
\textbf{Model} & \textbf{RTF} & \textbf{Pub WER} & \textbf{Priv WER} \\
\hline
Whisper-Medium (Tugstugi) & 0.182 & 0.44443 & 0.45064 \\
\hline
Wav2Vec2 & 0.045 & 0.51389 & 0.52338 \\
\hline
Titu-STT-BN-Conformer-Large & 0.005 & 0.44430 & 0.44550 \\
\hline
FT Whisper-Med (Adapter) & 0.019 & 0.31073 & 0.32145 \\
\hline
FT Titu-STT-BN-Conformer-Large & 0.005 & 0.40662 & 0.41434 \\
\hline
FT Whisper-Med (Contextual) & 0.019 & 0.35578 & 0.38231 \\
\hline
ROVER Ensemble & 0.019 & 0.33770 & 0.35762 \\
\hline
FT Whisper-Med (Faster Whisper)* & 0.019 & 0.30842 & 0.31070 \\
\hline
FT Whisper-Med (Faster Whip. + Demucs) & 0.068 & 0.30354 & 0.30612 \\
\hline
FT Whisper-Med (WhisperX) & 0.019 & 0.30406 & 0.31479 \\
\hline
\end{tabular}
\label{tab:asr_benchmarks}
\end{center}
\end{table}

\begin{table}[htbp]
\caption{Speaker Diarization Benchmarks}
\begin{center}
\begin{tabular}{|l|c|c|c|c|c|}
\hline
\textbf{Model} & \textbf{Retrained} & \textbf{Post-Proc} & \textbf{Pub DER} & \textbf{Priv DER} & \textbf{RTF} \\
\hline
%0.28077

%0.27893

Diarizen & Yes (WavLM) & Yes & 0.28077 & 0.27893 & 0.020 \\
\hline
Pyannote 3.1 & Yes & Yes & 0.20283 & 0.28145 & 0.019 \\
\hline
Pyannote Comm-1 & Yes & Yes & 0.24522 & 0.26640 & 0.019 \\
\hline
ECAPA-TDNN & No & Yes & 0.25322 & 0.31640 & 0.012 \\
\hline
\end{tabular}
\label{tab:dia}
\end{center}
\end{table}

\section{Challenges and Limitations}

During the competition, we encountered several challenges and limitations that impacted the overall performance and experimentation process:

\subsection{Compute Resource Constraints}
Our best model was trained on a single L40S GPU with 48 GB VRAM, which was limited to only 5 free hours per month from Lightening AI. This constraint made extensive experimentation and hyperparameter tuning difficult, limiting our ability to fully optimize the model. More compute hours could have significantly improved model performance.

\subsection{Data-Related Challenges}
The dataset consisted of multiple sources with varying quality, including missing or partial transcriptions. Handling these inconsistencies required significant preprocessing effort. Additionally, audio diarization issues made it difficult to accurately separate speakers in multi-speaker samples.

\subsection{Model Limitations}
Pre-trained models such as Titu fastConformer and Moonshine were not fully optimized for Bangla features. Fine-tuning was required, and low-epoch fine-tuning sometimes led to a decrease in performance. Overfitting was observed on smaller or incomplete subsets of the dataset.

\subsection{Evaluation Challenges}
Comparing different models was non-trivial due to differences in dataset coverage, preprocessing methods, and evaluation splits. Limited test set size further affected the reliability of evaluation metrics.

\subsection{Time and Experimentation Constraints}
The combination of limited GPU hours, large model sizes, and time-consuming preprocessing constrained the number of experiments we could conduct. This made it challenging to systematically explore model architectures, hyperparameters, and training strategies to achieve optimal performance.

\section{Conclusion}
This study highlights a critical dichotomy in low-resource speech processing. For Bengali ASR, raw data scaling and ensembling techniques are inferior to targeted fine-tuning on noisy audio paired with immaculate textual alignment. Conversely, for Bengali speaker diarization, state-of-the-art architectures and extensive fine-tuning fail to overcome the inherent acoustic complexities of long-form audio; instead, rigid algorithmic post-processing is the most effective method for minimizing the Diarization Error Rate. Our optimized pipelines, achieving $\sim$0.019 RTF, alongside the open-source release of the 882-hour Lipi-Ghor dataset, provide a robust foundation for the future of conversational AI in Bengali.

\section*{Acknowledgment}
Team Villagers would like to thank the organizers of DL Sprint 4.0 and BUET CSE FEST 2026 for providing the computational resources and dataset that made this research possible.
\nocite{*}
\bibliographystyle{IEEEtran}
\bibliography{references}

\end{document}